\documentclass[11pt]{article}
\usepackage{graphicx,xspace,amsmath,changebar,url,wasysym}
\usepackage{amssymb}
\usepackage[square,numbers]{natbib}
\usepackage[T1]{fontenc}
\usepackage[latin1]{inputenc}
\usepackage[english]{babel}
\newcommand{\ie}{\emph{i.e.}\xspace}

\newcommand{\be}{\begin{equation}}
\newcommand{\bel}[1]{\begin{equation}\label{#1}}
\newcommand{\ee}{\end{equation}}

\renewcommand{\H}{\mathcal{H}}
\newcommand{\Hp}{\H_{\text{p}}}

\newcommand{\LR}{\text{LR}}

\newcommand{\KinMix}{{\tt KinMix}}

\begin{document}
\title{{\sc Casework 
		applications of probabilistic genotyping methods for DNA mixtures that 
		allow relationships between contributors}}
\author{
Peter J. Green%\thanks {School of Mathematics, University of Bristol, Bristol BS8 1TW, UK. \newline \hspace*{5mm} Email: {\tt P.J.Green@bristol.ac.uk}.}
\\
UTS, Sydney, Australia\\University of Bristol, UK.\\
\and Julia Mortera%\thanks {Universit\`a Roma Tre, Italy.\newline \hspace*{5mm} Email: {\tt julia.mortera@uniroma3.it}}
\\
Universit\`a Roma Tre, Italy.\\
University of Bristol, UK.\\
\and
Lourdes Prieto
\\
Forensic Sciences Institute\\
 University of Santiago de Compostela, Spain\\
Comisar\'{i}a General de Polic\'{i}a Cient\'{i}fica, DNA Laboratory, Madrid, Spain}
\date{\today}
\maketitle
\begin{abstract}

 \end{abstract}
    \hspace{5mm}
In both criminal cases and civil cases there is an increasing demand for the analysis of DNA
mixtures involving relationships. The goal might be, for example, to identify the contributors to a DNA mixture where the unknown donors may be related, or to infer the relationship between individuals based on a DNA mixture.
This paper applies  a new approach to modelling and computation for DNA mixtures involving contributors with arbitrarily complex relationships to two real cases from the Spanish Forensic Police.

\noindent {\small {\em Some key words:} Coancestry, deconvolution, disputed relationship, identity by descent, kinship, DNA mixtures, likelihood ratio.}
 
\section{Introduction}
\label{sec:intro}

In both criminal  and civil cases relying on inference about relationships there is an increasing demand for the analysis of DNA mixtures where relatives are involved. The goal might be to identify the unknown contributors to a mixture where the donors may or may not be related, or  to determine relationships between typed individuals and one (or more) of the contributors to a mixture, also in the case that the mixture contributors themselves are related. Here we use a novel approach that is able to tackle these problems, which to our knowledge have not previously been analysed rigorously in the literature.  A new general software \texttt{KinMix} \texttt{R} package \citep{kinmix} which can handle complex relationships with and between mixture contributors has been developed to make inference in these cases. Inference is not limited to two-way relationships but can  be extended to relationships among 3 (or possibly more)  contributors to a mixture.

We analyse two real cases from the Spanish Forensic Police.  In the first case we wish to identify a missing person through the analysis of DNA mixtures found on personal belongings. In many cases, the genetic profile detected on the objects is not from a single source, but might be a DNA mixture, revealing that the object was used by 2 (or more) people. In addition, very often, the contributors to these mixtures are related, mainly in cases, such as this one,  where the missing person shared the dwelling with relatives.
 Here, among other analyses, we tackle the novel problem of computing a likelihood ratio that the two unknown contributors to the mixture are related compared to unrelated, testing  relationships such as parent-child, sibs, first cousins, etc.

The second case concerns a murder  where a man was stabbed in his home. A DNA  sample was taken from the murder weapon  and  appeared to be a  DNA mixture from the victim and possibly  a close relative of the victim.

 Here we use  probabilistic genotyping methods for DNA mixtures, under hypotheses about the  relationships among contributors to the mixture and to other individuals whose genotype is available.   We now briefly summarise these methods and refer to \citep{mortera2019dna} which presents a review on DNA mixtures where further background can be found.

A natural basis for any model-based {continuous} DNA mixture analysis is a joint model for the peak heights $\mathbf{z}$ in the electropherogram (EPG) and  genotypes $\mathbf{n}$,  $p(\mathbf{n},\mathbf{z}|\psi)= p(\mathbf{n}) \times p(\mathbf{z}|\mathbf{n},\psi)$, with parameters $\psi$ characterising the conditional distribution of peak heights \citep{graversen:package:13}. We base our analysis of DNA mixtures on the model for $p(\mathbf{z}|\mathbf{n},\psi)$ described in \citep{cowell:etal:15}. This model takes fully into account the variation in peak heights and the possible  artefacts, like stutter and dropout, that might occur in the DNA amplification process. The model can coherently analyse  a combination of replicates, a combinations of different
samples and a  combinations of different kits. 
%We refer to the review on DNA mixtures by  \citep{mortera2019dna}  for further details. 

In the standard case, unknown contributors to the mixture are assumed drawn at random from the gene pool. When the contributors are related, there is positive association between their contributor genotypes. A new model aimed at making inference about complex relationships from DNA mixtures is  presented in  \citep{green2020inference}. This generalises the work in  \citep{green:etal:17} which allowed inference about particular close relationships between contributors to a DNA mixture with unknown genotype and other individuals of known genotype.  
The new model  extends the analysis to different scenarios and allows to specify arbitrary relationships between a set of actors, each of which may be mixture contributors, or have measured genotypes, or both. We can evaluate the likelihood of any such model, and compare models accordingly. A brief description of the key ideas underlying, specifying, modelling and computing relationship inference is given in the Appendix.

%\begin{description}
%	
%	\item[(i)] we  have  the genotypes of some individuals that might be related to contributors to a mixture who are unrelated to each other;
%	\item[(ii)] we do not have any measured genotypes and  the contributors to one or more mixtures might be related to each other; 
%	\item[(iii)] we have some  measured genotypes and  the contributors to one or more mixtures might be related to each other and/or to some measured genotypes;
%\end{description}
%
%In all cases were one unknown contributor to the mixture is higher up the relationship pedigree than the other contributor   we need to test  whether the major unknown contributor $U_1$ is a descendant or an ancestor of $U_2$ -- the minor unknown contributor. 

The case work examples in  Section \ref{sec:Lourdes} illustrate some  scenarios,  where we make inference about  two-way relationships between two mixture contributors with and without information about their or their relatives' genotypes.

The software used to analyse the case work examples  is the new \texttt{KinMix} \texttt{R} package \citep{kinmix} that extends the \texttt{DNAmixtures} \texttt{R} package  \citep{graversen:package:13} to allow for modelling DNA mixtures with related contributors. 

Among existing published work on relationships and mixtures, \citep{slooten2018}  presents an empirical study with data known to include known sibs among the reference samples, used to broaden the basis for evaluation of the information gain from using peak height data.
Free software is also available to deal with DNA mixtures where contributors can be related \citep{bleka2016euroformix}, but this addresses a different problem: a specific kinship relationship has to be defined and one of the contributors has to be known. 

\section{Results of the analysis of complex DNA mixtures involving  relationship testing}
\label{sec:Lourdes}

In this section we demonstrate the results and performance of our methods on the two case studies.  For the first example we used the  data  gathered on 21 markers included in the GlobalFiler\textsuperscript{TM} Amplification kit (ThermoFisher) and in the second example we also used data on 16 markers in the PowerPlex\textsuperscript \textregistered 16 kit. In all examples we assume known  allele frequencies
 and adopt a threshold of 50 rfus.

\subsection{Example 1: Identification using personal belongings of a missing person}
\paragraph{Background on the case}

 Personal belongings such as toothbrushes or razor blades can be used as a source of DNA in missing person cases. In these objects, DNA from the missing person may be found since they may have been frequently used before his/her disappearance. Nevertheless, there is  uncertainty about the actual donor of the DNA isolated from these objects, and this is why it is recommended to ``validate'' the detected profile by using a reference (known) sample from a relative of the missing person. Usually, these profiles (from objects and/or relatives) are then compared with DNA profiles of unidentified bodies that are stored in national databases (``massive comparison''). This is useful to know if the missing person has passed away but his body was not identified. 
Unfortunately, in some cases, the genetic profile detected on an object is not a single source profile but a DNA mixture, revealing that the object was used by 2 (or more) people. In addition, very often, the contributors to these mixtures are related (mainly in cases where the missing person shared the dwelling with relatives). 

In this example, we present a real case related to a missing male. The full anonymised data together with the \texttt{R} scripts to compute the results are given in the Supplementary Material webpages\footnote{\texttt{https://petergreenweb.wordpress.com/example-1-data-code-and-output/}
}. The data are anonymised to avoid serious privacy and confidentiality concerns. 
 In this  case, only a daughter of the missing male was available to donate a DNA sample. This is not the ideal situation since false DNA matches can be found after a massive comparison of profiles in a database when only one relative is available as a reference sample. In order to improve the reference genetic data, a toothbrush and a razor-blade, presumably used by the missing person, were also collected. DNA from both objects was recovered and analysed by using the GlobalFiler kit (Thermo Fisher). The reference sample from the daughter of the missing male was also genotyped with the same kit. Two different DNA mixtures were detected in the two objects. An excerpt of the (anonymised) data is shown in Table 1, showing the alleles and peak heights in the two DNA mixtures found on the toothbrush T and the razor-blade RB. The DNA profile of the daughter, denoted by D, is also shown.
The sex-related markers indicated that the mixture was most probably from one female and one male contributor.

\paragraph{Results}
Here we analyse the two DNA mixtures found on the toothbrush T, and a razor-blade RB,  presumably  used by the missing person (ante-mortem data).

\begin{table}[htb]
	\centering
	\caption{Example 1: An excerpt of the anonymised data from the toothbrush $T$  and the razorblade $RB$, showing the markers, alleles and relative peak heights. The DNA profile of the daughter D of the missing person is also shown.}
	\label{tab:data}
	\vspace{2mm}
\begin{tabular}{l|rrrr}
\hline  	
&       alleles    &toothbrush             & razorblade &   \\
markers &   in mixture  & peak height & peak height &   D \\
\hline
Marker 6  &         17 &            &       945 &          \\
          &         19 &        264 &      853  &         19 \\
          &         21 &        3664 &       612 &         21 \\
Marker 7  &         13  &       1152 &       245&            \\
          &         14 &         126 &       796 &          \\
          &15          &         941 &       830 &        15 \\
Marker 14 &         13 &       5158 &       2141 &         13 \\
          &         15 &        304 &       1512 &         15 \\
Marker 20 &         13 &       3218 &        334 &            \\
          &         17 &       3550 &       1795 &         17 \\
          &         18 &            &       1274 &            \\
					\hline
\end{tabular}  
\end{table}

\begin{table}[htb]
  \centering
   \caption{Example 1: Estimated parameters based on an analysis of the two mixture samples assuming that the toothbrush $T$ and $RB$ contain DNA from two unknown contributors.}
  \label{tab:par}
 	\vspace{2mm}
\begin{tabular}{r|ccccc}
 	\hline
           &     $\mu$ &  $\sigma$ &       $\xi$ &     $\phi_{U_1}$ &    $\phi_{U_2}$ \\
					\hline
toothbrush		&      2381  &       0.0614 &     0 &      0.9262    & 0.0736       \\
razor-blade &      1602  &   0.0504   &  0.0118  & 0.5001 & 0.4999    \\  
\hline
\end{tabular}  
\end{table}
 
\begin{table}[htb]
  \centering
   \caption{Example 1: $\log_{10} LR$ for testing whether in $T$  and $RB$, $\H_p$ contributor ($U_1$ or $U_2$) is a parent of  D \textit{vs.} $\H_0$ no contributor is related to D.}
  \label{tab:test}
 	\vspace{2mm}
\begin{tabular}{r|r|r}
 	\hline
         &  \multicolumn{2}{c}{$\log_{10} \LR$}      \\
          & {$U_1$}    & {$U_2$}   \\
	\hline
toothbrush		&       10.974     & 4.531     \\
razor-blade   &           8.443   &  8.444  \\  
\hline
\end{tabular}  
\end{table}
 Table \ref{tab:par} shows the estimated parameters $\psi=(\mu,\sigma,\xi,\mathbf{\phi})$ for the analysis of the DNA mixtures found on $T$ and $RB$. We assume there are 2 unknown contributors, denoted $U_1$ and $U_2$, to each of $T$  and $RB$: not necessarily the same individuals in the two cases. We fix on two contributors since the analysis performed for 3 (not shown here) yielded an almost vanishing proportion for the third contributor, $\phi_3 =0$. Note also that the stutter proportion $\xi$ for sample $T$ is zero indicating that stutter peaks were most probably removed from the data. The estimated proportion of DNA for the two contributors to sample $T$ is large for the major contributor $U_1$, $\phi_{U_1}=0.93$, whereas, for item   $RB$  the estimated proportions of DNA contributed by $U_1$ and $U_2$  are roughly equal, 
$\phi_{U_1} \simeq \phi_{U_2}= 0.5$,  implying they contributed in almost equal proportions to the mixture. 
 As we will see in the latter case the estimation of the \LR\ and other inference is problematic.  In these models, the likelihood can have a complicated shape and numerical maximisation can be unreliable.  The values in Table \ref{tab:par} are  the maximum likelihood estimates as calculated by \texttt{DNAmixtures}.

Table \ref{tab:test} shows the  $\LR$ and $\log_{10} LR$ for testing  $\Hp$: D is the child of $U_1$ (and similarly for  $U_2$)  {\emph{vs.}} 
$\H_0$: no unknown contributors are related to D.  For item $T$, $\log_{10} \LR=10.97$ is large pointing to  $U_1$ being a parent of D. It is also substantial for the hypothesis concerning $U_2$ being a parent of D. Could this be due to the fact that the two contributors might be related?  We will test this assumption later. For the $RB$ the $\log_{10} \LR$ in Table \ref{tab:test} for  $\H_1$ \textit{vs.}  $\H_0$  is almost the same when testing whether D is the child of  $U_1$  or of $U_2$. This is probably due to the fact that the proportions are almost identical, $\phi_{U_1} \simeq \phi_{U_2}= 0.5$,  which makes it extremely difficult to distinguish between the contributors. 

We also tested  whether the daughter D was a contributor to $T$   or not, similarly for $RB$, and in both cases the log\LR\ was zero, excluding D from being a contributor to either mixture.

\begin{table}[htb]
	\centering
	\caption{Example 1: Excerpt 
		of marker-wise $\LR$ and overall $\log_{10}\LR$ for item $T$, using \texttt{relMix} and
		 \texttt{KinMix} with and without peak height information, for testing whether in $T$,  $\H_p$: $U_1$ is a parent of D \textit{vs.} $\H_0$: $U_1$ and $U_2$ are random members of the population.}
	\label{tab:RelKin}
	\vspace{2mm}
\begin{tabular}{l|rrr}
	\hline
marker	&     \texttt{relMix} &     \texttt{KinMix}&     \texttt{KinMix} \\
	   &                      &     w/o peak heights  &  with peak heights \\
\hline
Marker 6 &       2.55 &  2.58 &     3.34 \\
	Marker 7 &       1.08 &    1.07 &   1.59 \\
Marker 9 &       1.26 &  1.18 &     1.62 \\
	Marker 14 &       2.09 &  2.12 &     1.51 \\
	
	\hline
partial  	$\log_{10}\LR$ &       8.35 & 8.42 &     9.94 \\
overall 	$\log_{10}\LR$ &       & 9.53 &     10.97  \\	
\hline
\end{tabular}  
\end{table}

In Table \ref{tab:RelKin} we present comparisons with the results of another freely-available package that analyses DNA mixtures involving relatives, \texttt{relMix} \citep{hernandis2019relmix}; this uses allele-presence only, not peak heights. We compare, marker-by-marker, with \texttt{KinMix} both with and without peak height information.
The results obtained with \texttt{relMix} and \texttt{KinMix} when not including the peak height information (columns 2 and 3) are quite similar. Small  differences between \texttt{relMix} and \texttt{KinMix} when not including peak heights are to be expected since they are based on different statistical models for the mixture. For the majority of  markers, when including peak height information  \texttt{KinMix} gave a larger $\log_{10} \LR$ (10.97 compared with 9.53, corresponding to a \LR\ 27.5 times smaller). For two of the markers, Markers 8 and 10, \texttt{relMix} is unable to compute the likelihood, most likely because of excessive storage demands. We can compute ``partial'' $\log_{10}\LR$s by excluding these 2 markers, and these are also shown in the Table.

\begin{table}[htb]
  \centering
   \caption{Example 1: For items $T$ and $RB$, $\log_{10} \LR$ for $\Hp$: the two contributors to the mixture are related, \ie\ $U_1$ has relationship $R$ to $U_2$,   \textit{vs.} $\H_0$: the two contributors are unrelated. Several different relationships $R$ are tested.}
  \label{tab:TRBpedLR}
	\vspace{2mm}
 \begin{tabular}{r|rr}
 	\hline
Relationship $R$ between $U_1$ to      &  $T$ & $RB$ \\
 and $U_2$ under $\Hp$          & \multicolumn{2}{c}{$\log_{10} \LR$}     \\
\hline
monozygotic twins &  $- \infty$  & $-\infty$    \\
parent-child &  $- \infty$  & $-\infty$    \\
sibs &  $-2.14$ &   $-$2.85  \\
double first cousins & $-0.510$ & $-0.657$ \\
quadruple-half-first-cousins &  $-0.44$ &  $-$0.630 \\
half-sibs & $-0.37$ &$-$0.625   \\
first cousins & $-0.10$ &   $-$0.148    \\
half-cousins & $-0.034$ &  $-$0.037   \\
\hline
\end{tabular}  
\end{table}

Table \ref{tab:TRBpedLR} shows the results for testing whether the contributors $U_1$  or $U_2$ to item $T$ and $RB$ are related, \ie\  \emph{$\Hp$: $U_2$ has relationship $R$ to  $U_1$} \emph{versus}
\emph{$\H_0$: $U_1$ and $U_2$ are unrelated}. The $\log_{10}\LR$s are all negative, implying that the $\LR$s are smaller than 1. Although only a finite set of possible relationships has been considered, these vary widely, and it is overwhelmingly clear there is there is no support for any relationship between the two contributors.  

We now consider the toothbrush EPG in more detail, examining the joint relationships between the mixture contributors and the typed daughter D, which clarifies the role of D in validating the mixture profile. Table \ref{tab:Trel} shows the  $\log_{10} \LR$ for item $T$  for several hypotheses $\Hp$ concerning different relationships  $R$ among $U_1$, $U_2$ and D,  \textit{vs.} the null hypothesis that these individuals are all unrelated. The values of the $\log_{10} \LR$ show that there is strong evidence that the two contributors to item $T$ are the missing father of D and D's mother, or at least very close relatives of them. Comparing the first 4 rows of Table \ref{tab:Trel} confirms that the most likely single possibility is that they are indeed the mother and father. All values in the Table remain unchanged if the sexes of all contributors are reversed; we choose to identify them in the way shown because inference (not shown) also including the Amelogenin locus indicates that is is most likely that the major contributor $U_1$ is female.

If there is interest in comparing two of the models displayed in Table \ref{tab:Trel}, the appropriate $\log_{10} \LR$ is simply obtained by calculating the difference beteen the values shown. For example, comparing the first row and the fifth, $17.935-10.974 = 6.961$ gives the weight of evidence that $U_2$ is the father of D, given that it is already assumed that $U_1$ is the mother of D. There are too many different such comparisons that can be made to list them all here.

Some of the specific relationships examined in Table \ref{tab:Trel} are speculative, but might be of interest in cases where a home is shared by an extended family.

\begin{table}[htb]
	\centering
	\caption{Example 1: For item $T$, $\log_{10} \LR$ for several hypotheses $\Hp$ concerning different relationships  $R$ among $U_1$, $U_2$ and D,  \textit{vs.} $\H_0$: $U_1$ and $U_2$ and D are unrelated. The results in the lower half of the table can be used as baselines for comparison for those in the upper half. All $\log_{10} \LR$ remain unchanged if the sexes of $U_1$ and $U_2$  are switched.}
	\label{tab:Trel}
		\vspace{2mm}
\begin{tabular}{l|r}
		\hline
		\multicolumn{1}{c|}{$\H_p$} &     $\log_{10} \LR$        \\
		\hline
$U_1$ mother of D and $U_2$ father of D  &          17.935 \\
$U_1$ maternal aunt of D and $U_2$ father of D &          14.028 \\
$U_1$ mother of D and $U_2$ paternal uncle of D &         15.579\\
$U_1$ maternal aunt of D and $U_2$ paternal uncle of D &        11.763\\
\hline
$U_1$ mother of D and $U_2$ unrelated & 10.974    \\
$U_1$ maternal aunt of D and $U_2$ unrelated & 7.452    \\
$U_1$ unrelated and $U_2$ father of D &  4.530   \\
$U_1$ unrelated and $U_2$ paternal uncle of D  &  2.796   \\
\hline
\end{tabular}  
\end{table}

Finally for this example, we consider the two mixture profiles $T$ and $RB$ jointly. What is the strength of evidence that the same individuals have contributed to both mixtures, and if so, are they related to D? To answer such questions, we use \KinMix\ to model various scenarios which deal with the two DNA mixture traces simultaneously, with various patterns among the contributors. There are too many permutations to show them all, so in Table \ref{tab:joint} we just present some interesting examples. As parameters for these joint peak height model, we copy over the relevant values from Table \ref{tab:par}. For full details of these calculations, please consult the codes in the online Supplementary material.

Table \ref{tab:joint} shows strong support for the hypothesis 
that the contributors to $T$ and $RB$ overlap and are mostly likely identical, strengthened further when a common contributor is a parent to D. As in previous analyses, the results are unchanged when sexes are interchanged, and in each hypothesis concerning a parent, the possibility that it is a close relative instead could also be examined. 

\begin{table}[htb]
	\centering
	\caption{Example 1: $\log_{10} \LR$ for the joint analysis of several hypotheses  concerning the identity between contributors to $T$ and $RB$ and whether a common contributor is a parent of D. In all cases, the baseline $\H_0$ states that both contributors and D are unrelated. All $\log_{10} \LR$ remain unchanged if the sexes of the contributors are switched. In the last two rows, the contributors are mentioned in order, major then minor, omitted for brevity.}
	\label{tab:joint}
	\vspace{2mm}
\begin{tabular}{l|r}
	\hline
$\H_p$ &     $\log_{10} \LR$        \\
\hline
$T$ and $RB$ have same 2 contributors  &      23.56 \\
$T$ and $RB$ have same 2 contributors, first being parent of D  &      34.54 \\
$T$ and $RB$ have same major contributor  &      16.53 \\
$T$ and $RB$ have same major contributor, being parent of D  &      27.50 \\
$T$ has father and mother of D, $RB$ has father and unknown 
%$T$ has father and mother (major and minor contributors) of D, $RB$ has father and unknown
& 34.46 \\
$T$ has mother and father of D, $RB$ has father and unknown & 25.54 \\
\hline
\end{tabular}  
\end{table}

\subsection{Example 2: Analyses of a  Spanish murder case}

\paragraph{Description of the case}

This concerns a murder case where a man was stabbed in his home. There was a knife with blood at the crime scene. 
The blood was mainly on the blade, but there was also some blood on the handle. The sample from the handle turned out to be a DNA mixture, with a major profile matching the victim. We also wish to test whether the minor profile in the mixture could be  a close relative of the victim (possibly a son). The DNA profile of the son was not available. Two EPGs from the mixture were obtained by using two different kits, we denote these by EPG1 and EPG2. The kits have partially overlapping sets of markers,  EPG1 was analysed on  its 16 markers and  EPG2 on its set of 22 markers, both  include Amelogenin. Here we assume known allele frequencies taken from the Spanish allele frequency database collected on n = 284 individuals \citep{garcia2012}.

Months after the murder, a man was arrested for a different crime, drug trafficking, and a reference DNA sample was collected. When  his profile was entered in the  DNA database several matches were found, among which with the DNA mixture on the handle of the knife. The  matches were investigated and  the identity of the person (name, date of birth, place of birth, name of the father, name of the mother) was that of the son of the victim.
Table \ref{tab:data2} gives an excerpt of the data  showing the markers, alleles and relative peak heights for EPG1 and EPG2, together with the genotypes of the father (the victim) and the son (the suspect). 

\begin{table}[htb]
	\centering
	\caption{Example 2: An excerpt of the data showing the markers, alleles and relative peak heights for EPG1 and EPG2, together with the father's and son's genotypes}
	\label{tab:data2}
	\vspace{2mm}
\begin{tabular}{r|rrrrr}
	\hline	
		&            &       EPG1 &       EPG2 &    father  &    son \\
	marker &     allele &     height &     height &            &            \\
\hline
	CSF1PO &         10 &        305 &        625 &         10 &         10 \\
 &         11 &        240 &        504 &         11 &         11 \\
	D10S1248 &         13 &            &       6990 &         13 &            \\
	 &         14 &            &       2309 &            &         14 \\
	 &         16 &            &       7144 &         16 &         16 \\
	D7S820 &          9 &        606 &       1136 &          9 &          9 \\
	       &         10   &            &        686 &         10 &            \\
	TH01 &        9.3 &        863 &       2654 &        9.3 &        9.3 \\
	 &         10 &        570 &            &            &         10 \\
	\hline
\end{tabular}  
\end{table}

\paragraph{Results}
We analysed the data from this case to illustrate the different scenarios that can be analysed using the recently developed  \texttt{Kinmix} code.

In particular we analyse the following different possible scenarios: 

\begin{description}
	\item[Scenario  1] Here none of the contributors are typed. The 
 analysis is of a 2-person mixture model for a prosecution hypothesis $\H_p:$ being the two unknowns being father and son versus $\H_0$ the two unknown contributors are unrelated.

\item[Scenario  2] Here only the  father (the victim) is typed. The analysis is of a 
2-person mixture model, where father has been typed and the prosecution hypothesis is $\H_p:$ son of father and 1 unknown are contributors versus $\H_0:$ no contributor is related to the typed individual (the victim). 

\item[Scenario  3] Both father and son are typed. Here we analyse a  2-person mixture model where $\H_p$: the contributors are victim (father) and son versus $\H_0:$ contributors to the mixture(s) are 2 unknown individuals.

\item[Scenario  4]
Both father and son are typed. Here we analyse  a  2-person mixture model where $\H_p$: the contributors are victim (father) and son versus $\H_0:$ contributors to the mixture(s) are the victim and an  unknown.
\end{description}

In all scenarios, unless otherwise stated, when considering an unknown contributor  to a mixture,  he or she is taken to be a random member of the reference population, so unrelated to typed individuals.

For EPG1 the  MLEs of the parameters under both $\H_p$ and $\H_0$ are similar and are roughly equal to $\psi=(\mu=576, \sigma=0.32, \xi=0, \phi_{U_1}=0.88,\phi_{U_2}=0.12)$. When the victim's genotype is known the estimated proportion contributed to EPG1 is $\phi_{v}=0.18$, $\phi_{U_1}=0.82$. For EPG2  the  MLEs of the parameters are roughly equal to $\psi=(\mu=2542, \sigma=0.97, \xi=0, \phi_{U_1}=0.75,\phi_{U_2}=0.25)$.  When the victim's genotype is known the estimated proportion contributed to EPG1 is $\phi_{v}=0.14$, $\phi_{U_1}=0.86$. In both EPG1 and EPG2 the victim is estimated to be the minor contributor.  Note that EPG2 has a higher  $\mu$ than EPG1 but this is also accompanied by a larger $\sigma$, so the coefficient of variation is similar in both EPGs. The MLEs of the mean stutter proportion $\xi$ are zero, probably because preprocessing of the data has removed peaks that were classified in the laboratory as stutter. Our models, however, allow for stutter and do not require that the data be preprocessed before analysis.

Table \ref{tab:Cases} gives the $\log_{10}\LR$ for the 4 scenarios when analysing EPG1 and EPG2 separately and jointly.  When combining EPGs made from the same DNA extract, as in this case, it is natural to make an
assumption that contributors are the same.  In \citep{graversen2019yara} we show how  results based on a combination of replicates, a combinations of different
samples and a combinations of different kits improve the robustness of  the analysis and help in fixing  any complications relating to degradation. However, when combining
profiles from different samples one needs to carefully consider
whether there is perhaps only a partial overlap.

\begin{table}[htb]
	\centering
	\caption{Example 2: $\log_{10}\LR$ for Scenarios 1--4 using EPG1 and EPG2 separately and in combination.}
	\label{tab:Cases}
		\vspace{2mm}
\begin{tabular}{r|rrrr}
		\hline
		Scenario   &          1 &          2 &          3 &          4 \\
		Typed actors &      \multicolumn{1}{c}{none} &     \multicolumn{1}{c}{father}  & \multicolumn{2}{c}{father \& son}  \\
		\hline
		EPG1&        $-$0.806 &       5.60 &      22.16 &      22.78 \\
		EPG2 &  	   $-$0.175 &      10.66 &      29.16 &      11.68 \\
		EPG1 \& EPG2 & 2.49  &  	  8.26 &	  40.17	&      26.20  \\
		\hline
	\end{tabular}  
\end{table}

\begin{table}[htb]
	\centering
	\caption{Example 2: For item EPG1 and EPG2, $\log_{10} \LR$ for $\Hp$: the two contributors to the mixture are related, \ie\ $U_1$ has relationship $R$ to $U_2$,   \textit{vs.} $\H_0$: the two contributors $U_1$ and $U_2$ are unrelated and are independent of the typed individuals. Several different relationships $R$ are tested.}
	\label{tab:rel2}
	\vspace{2mm}
\begin{tabular}{r|rr}
	\hline
	& \multicolumn{2}{c}{$\log_{10} \LR$}\\
Relationship	&       EPG1 &       EPG2 \\
		\hline
	parent-child  &     $-$0.806 &     $-$0.175 \\
	sibs  &     $-$1.270 &     $-$0.940 \\
	quadruple-half-first-cousins  &     $-$0.316 &     $-$0.022 \\
	half-sibs  &     $-$0.275 &      0.045 \\
	first cousins  &     $-$0.108 &      0.059 \\
	half-cousins  &     $-$0.046 &      0.040 \\
	\hline
\end{tabular}  
\end{table}

Table \ref{tab:rel2} shows $\log_{10} \LR$ for testing whether the two unknown contributors to the DNA mixture are related  versus that they are unrelated.  For EPG1 the $\LR$s for testing $\Hp$ that the $U_1$ has a relationship $R$ to $U_2$,   \textit{vs.} $\H_0$: the two contributors $U_1$ and $U_2$ are unrelated  and are independent of the typed individuals,  vary between 0.16 and 0.9 giving roughly equal weight to $\H_1$ versus $\H_0$. For   EPG2 these vary between 0.11 and 0.86.

%\begin{table}[htb]
%	\centering
%	\caption{Estimated parameters for the joint analysis of the different scenarios concerning different  hypotheses and typed individuals for Cases 1-4 using mixtures 1 and 2.}
%	\label{tab:par3}
%	\begin{tabular}{r|rrrrrrrrrr}
%		Hypotheses    & $\mu$ & $\sigma$ & $\xi$ & $\phi_{a}$ &$\phi_{b}$& $\mu$ & $\sigma$ & $\xi$ & $\phi_{a}$ &$\phi_{b}$	\\  
%		$\H_0:U_1\&U_2$ &	567&    0.530 &    2.8$\times10^{-08}$ &      0.556 &      0.444 &       2558 &      0.977 &   8.9$\times10^{-22}$ &      0.396 &      0.605 \\
%		$Hd: v\&U_1$ &	575 &      0.305 &     0.0182 &      0.819 &      0.181 &       2541 &      0.749 &   1.15$\times10^{-12}$ &      0.207 &      0.793 \\
%		$U_1, U_2$  Scenario 1  &	567 &      0.530 &   7.5$\times10^{-09}$ &      0.538 &      0.462 &       2558 &      0.977 &   1.9$\times10^{-21}$ &      0.807 &      0.193 \\
%		$U_1, U_2$  Scenario 2   &	567 &      0.530 &    2.8$\times10^{-08}$ &      0.444 &      0.556 &       2558 &      0.977 &   1.7$\times10^{-22}$ &      0.605 &      0.396 \\
%		$v\&s$ Scenario 3 & 	567 &      0.530 &     0.0123 &      0.656 &      0.344 &       2558 &      0.977 &   6.7$\times10^{-23}$ &      0.338 &      0.662 \\
%	\end{tabular}  
%\end{table}

Table \ref{tab:decon} shows the deconvolution for the major contributor to the mixture for the two EPGs. The table only
indicates genotype probabilities of at least 0.001, meaning that cells with a probability of less than 0.001 have been suppressed. We have  denote by \textit{other} the collection of alleles for which no peak has been observed in the EPG. For EPG1 the highest ranking genotype for the major contributor $U_1$ on all markers has  posterior probability greater than 0.99 and coincides with the genotype of the suspect (who is the son of the victim) on all markers. The deconvolution for  EPG2  gives a much poorer performance.  For example, on marker D7S850 the top ranking genotype for EPG2 is incorrect, the correct genotype (9,9) is ranked 3$^{\mbox{rd}}$ having a small  probability of 0.077.

\begin{table}[htb]
	\centering
	\caption{Example 2: Predicted genotypes of $U_1$ with corresponding  probabilities for  EPG1 and EPG2  for an excerpt of the markers.  An allele not observed in the EPG is
	 denoted by \textit{other}.}
	\label{tab:decon}
		\vspace{2mm}
\begin{tabular}{r|ccr|ccr}
		\hline
		&      \multicolumn{3}{c|}{EPG1}              &     \multicolumn{3}{c}{EPG2}                \\
	& \multicolumn{2}{c}{genotype} & prob. & \multicolumn{2}{c}{genotype} & prob. \\
	\hline
	CSF1PO &  10 &         11 &     1      &         10 &         11 &   0.751 \\
	&            &            &            &         10 &         10 &   0.097 \\
	&            &            &            &         11 &         11 &   0.083 \\
	&            &            &            &         10 &         \textit{other} &   0.036 \\
	&            &            &            &         11 &         \textit{other} &   0.033 \\
		&            &            &            &            &            &            \\
	D13S317 &         12 &         13 &      0.997 &         12 &         13 &   0.576 \\
	&         12 &         12 &      0.003 &         12 &         12 &   0.363 \\
	&            &            &            &         12 &         \textit{other} &   0.043 \\
	&            &            &            &         13 &         \textit{other} &   0.011 \\
	&            &            &            &         13 &         13 &   0.006 \\
	&            &            &            &            &            &            \\
	D7S820 &   9 &          9 &      1     &          9 &         10 &   0.768 \\
	&            &            &            &         10 &         10 &   0.077 \\
	&            &            &            &          9 &          9 &    0.077 \\
	&            &            &            &          9 &         \textit{other} &   0.043 \\
	&            &            &            &     5   10 &         \textit{other} &   0.034 \\
	&            &            &            &            &            &            \\
	TH01 &   9.3 &        10  &   1   &        9.3 &        9.3 &    0.812 \\
	&            &            &            &        9.3 &         \textit{other} &   0.185 \\
	&            &            &            &         \textit{other} &         \textit{other} &   0.003 \\
%	TPOX &     8 &       9    &   0.999    &          8 &         11 &   0.715 \\
%	&          8 &          8 &   0.001    &          8 &          8 &   0.010 \\
%	&            &            &            &         11 &         11 &   0.083 \\
%	&            &            &            &         11 &         \textit{other} &   0.054 \\
%	&            &            &            &          8 &         \textit{other} &   0.048\\
\hline
\end{tabular}  
\end{table}
\clearpage

\section{Conclusions} 

We have shown that a wide range of relationship inference problems where one or more actors appear only
as contributors to a DNA mixture, can be handled coherently.  We can make inference about relationships among contributors, and between contributors and typed individuals. We  carried out diagnostic plotting (not shown here) as recommended by \cite{graversen2019yara} 
and found nothing to suggest the model was failing to fit the data.

The new \texttt{KinMix} package \citep{kinmix} used  in the casework examples illustrated here is a highly flexible modular software package capable of solving much more complex relationships among two or more mixture contributors than those presented here. It is not limited to pairwise relationships. In \citep{green2020inference} we show its capabilities of dealing with multi-way relationships in DNA mixtures including cases where the contributors might be inbred. 

\newpage

\section*{Appendix} 
 
 The key idea that enables the specification, modelling and computation of DNA mixtures with familial relationships among the contributors, and/or between contributors and other typed individuals is the IBD pattern distribution. IBD stands for identity by descent, the phenomenon where two or more related individuals have a common allelic value at a marker, not by the coincidence of several draws from the gene pool giving the same value, but because the allele was passed from parent to child in the process of meiosis. For a given set of related individuals, or 'family', an IBD pattern is a partition of the alleles of the individuals in the family according to their identity by descent. The IBD pattern distribution is simply the probability distribution of this partition induced by repeated application of Mendel's first law.
 
 For just two related contributors, the idea  has been in use to quantify relatedness for 80 years, in the form of Cotterham's kappas; for example, the relationship between two full brothers is captured by the probabilities that 0, 1 or 2 alleles are identical by descent: $\kappa_0=0.25$, $\kappa_1=0.5$, $\kappa_2=0.25$. The IBD pattern distribution extends this notion to any number of related individuals, and also deals correctly with inbreeding.
 
In \texttt{KinMix}, the IBD pattern distribution is used not only to specify the relationships in question, but also to model the distribution of the genotype profiles, and as a data structure to drive the computation. With unlinked autosomal STR markers in Hardy-Weinberg equilibrium, the joint distribution of the genotype profiles of the family members is completely determined by the IBD pattern distribution and the allele frequencies for each marker. As in much other recent work on computation for STR probabilistic genotyping methods for mixtures, joint distributions of genotype profiles are implemented using Bayesian networks (BNs), which allow efficient exact computation. The IBD pattern distribution is used directly in building the BN for the related genotypes. Full details are given in \cite{green2020inference}, and the methodology is implemented in the \texttt{R} package \texttt{KinMix} \citep{kinmix}.
  
%\bibliographystyle{abbrvnat}
%\bibliographystyle{plainnat}
%\bibliography{dna1}

\begin{thebibliography}{11}
\providecommand{\natexlab}[1]{#1}
\providecommand{\url}[1]{\texttt{#1}}
\expandafter\ifx\csname urlstyle\endcsname\relax
  \providecommand{\doi}[1]{doi: #1}\else
  \providecommand{\doi}{doi: \begingroup \urlstyle{rm}\Url}\fi

\bibitem[Bleka et~al.(2016)Bleka, Storvik, and Gill]{bleka2016euroformix}
{\O}.~Bleka, G.~Storvik, and P.~Gill.
\newblock Euro{F}or{M}ix: an open source software based on a continuous model
  to evaluate str dna profiles from a mixture of contributors with artefacts.
\newblock \emph{Forensic Science International: Genetics}, 21:\penalty0 35--44,
  2016.

\bibitem[Cowell et~al.(2015)Cowell, Graversen, Lauritzen, and
  Mortera]{cowell:etal:15}
R.~G. Cowell, T.~Graversen, S.~L. Lauritzen, and J.~Mortera.
\newblock Analysis of {DNA} mixtures with artefacts (with discussion).
\newblock \emph{Journal of the Royal Statistical Society: Series C},
  64:\penalty0 1--48, 2015.

\bibitem[Garc\'{i}a et~al.(2012)Garc\'{i}a, Alonso, Cano, Garc\'{i}a, Luque,
  Mart\'{i}n, de~Yuso, Maulini, Parra, and Yurrebaso]{garcia2012}
O.~Garc\'{i}a, J.~Alonso, J.~A. Cano, R.~Garc\'{i}a, G.~M. Luque,
  P.~Mart\'{i}n, I.~M. de~Yuso, S.~Maulini, D.~Parra, and I.~Yurrebaso.
\newblock Population genetic data and concordance study for the kits
  {I}dentifiler, {NGM}, {P}ower{P}lex {ESX} 17 {S}ystem and {I}nvestigator
  {ESS}plex in {S}pain.
\newblock \emph{Forensic Science International: Genetics}, 6\penalty0
  (2):\penalty0 e78--e79, 2012.

\bibitem[Graversen(2013)]{graversen:package:13}
T.~Graversen.
\newblock \emph{{DNAmixtures}: Statistical Inference for Mixed Traces of
  {DNA}}, 2013.
\newblock {R} package version 0.1-4.
  \url{http://dnamixtures.r-forge.r-project.org}.

\bibitem[Graversen et~al.(2019)Graversen, Mortera, and Lago]{graversen2019yara}
T.~Graversen, J.~Mortera, and G.~Lago.
\newblock The {Y}ara {G}ambirasio case: Combining evidence in a complex {DNA}
  mixture case.
\newblock \emph{Forensic Science International: Genetics}, 40:\penalty0 52--63,
  2019.

\bibitem[Green(2020)]{kinmix}
P.~J. Green.
\newblock \emph{KinMix: {DNA} mixture analysis with related contributors},
  2020.
\newblock R package version 2.0,
  \texttt{https://petergreenweb.wordpress.com/kinmix2-0}.

\bibitem[Green and Mortera(2017)]{green:etal:17}
P.~J. Green and J.~Mortera.
\newblock Paternity testing and other inference about relationships from {DNA}
  mixtures.
\newblock \emph{Forensic Science International: Genetics}, 28:\penalty0
  128--137, 2017.
\newblock {\tt http://dx.doi.org/10.1016/j.fsigen.2017.02.001}.

\bibitem[Green and Mortera(2020)]{green2020inference}
P.~J. Green and J.~Mortera.
\newblock Inference about complex relationships using peak height data from
  {DNA} mixtures, 2020.
\newblock {\tt https://arxiv.org/abs/2005.09365}.

\bibitem[Hernandis et~al.(2019)Hernandis, D{\o}rum, and
  Egeland]{hernandis2019relmix}
E.~Hernandis, G.~D{\o}rum, and T.~Egeland.
\newblock rel{M}ix: An open source software for {DNA} mixtures with related
  contributors.
\newblock \emph{Forensic Science International: Genetics Supplement Series},
  2019.
\newblock \texttt{https://doi.org/10.1016/j.fsigss.2019.09.085}.

\bibitem[Mortera(2020)]{mortera2019dna}
J.~Mortera.
\newblock {DNA} mixtures in forensic investigations: The statistical state of
  the art.
\newblock \emph{Annual Review of Statistics and Its Application}, 7:\penalty0
  1--34, 2020.
\newblock \texttt{https://doi.org/10.1146/annurev-statistics-031219-041306}.

\bibitem[Slooten(2018)]{slooten2018}
K.~Slooten.
\newblock The information gain from peak height data in {DNA} mixtures.
\newblock \emph{Forensic Science International: Genetics}, 36:\penalty0
  119--123, 2018.

\end{thebibliography}

\end{document}